\documentclass[12pt,a4paper]{article}
\pdfoutput=1
\usepackage[utf8]{inputenc}
\usepackage[T1]{fontenc}
\usepackage[margin=1.9cm]{geometry}
\usepackage[pdfstartview={FitH},bookmarks=false,linktoc=page,colorlinks=true,linkcolor=blue,citecolor=blue,urlcolor=blue]{hyperref}
\usepackage[numbered]{bookmark}
\usepackage{mathtools}
\usepackage{amssymb}
\usepackage{graphicx}
\usepackage[font=small,labelfont=bf]{caption}
\usepackage{authblk}
\usepackage{cite}
\usepackage{dsfont}
\usepackage[dvipsnames]{xcolor}
\usepackage{xcolor}
\usepackage[titles]{tocloft}
\usepackage{float}
\usepackage{subcaption}
\usepackage{color}
\usepackage{tikz}
\usepackage{ifthen}
\usepackage[warn]{textcomp}
\numberwithin{equation}{section}
\allowdisplaybreaks
\usepackage{multirow}
\usepackage{bbold}
\usepackage[toc,page]{appendix}

\usetikzlibrary{patterns}

\newcommand{\ket}[1]{|#1\rangle}
\newcommand{\bra}[1]{\langle#1|}

\usepackage{comment}

\usepackage{pgfplots}
\pgfplotsset{compat=1.9}

\usetikzlibrary{arrows.meta, patterns, intersections, backgrounds}
\usepgfplotslibrary{fillbetween}

\tikzset{ntp/.style={circle, thin, minimum size=2mm, inner sep=0,
fill=white,#1}}

\usepackage{caption}
\usepackage{subcaption}

\begin{document}
\title{\vspace{2cm}\textbf{ Complexity of Quantum Harmonic Oscillator in External Magnetic Field }\vspace{1cm}}

\author[a]{V. Avramov}
\author[a]{M. Radomirov}
\author[a,b]{R. C. Rashkov}
\author[a]{T. Vetsov}

\affil[a]{\textit{Department of Physics, Sofia University,}\authorcr\textit{5 J. Bourchier Blvd., 1164 Sofia, Bulgaria}
	
\vspace{-10pt}\texttt{}
\vspace{0.2cm}}

\affil[b]{\textit{Institute for Theoretical Physics, Vienna University of Technology,}
  \authorcr\textit{Wiedner Hauptstr. 8–10, 1040 Vienna, Austria}
  
\vspace{20pt}\texttt{v.avramov,radomirov,rash,vetsov@phys.uni-sofia.bg}\vspace{0.1cm}}
\date{}
\maketitle

\begin{abstract}
In this paper, we investigate the circuit complexity of a quantum harmonic oscillator subjected to an external magnetic field. Utilizing the Nielsen approach within the thermofield dynamics (TFD) framework, we determine the complexity of thermofield double states as functions of time, temperature, and the external magnetic field. Our subsequent analysis reveals various features of this complexity. For instance, as temperature increases, the amplitude of complexity oscillations also rises, while at low temperatures, complexity stabilizes at a constant positive value. Furthermore, the magnetic field creates two distinct sectors: strong magnetic fields exhibit periodic complexity oscillations, whereas weak magnetic fields induce a beating effect. Finally, we confirm that the rate of complexity obeys the Lloyd bound.
\end{abstract}

\tableofcontents

\section{Introduction}

In recent years, there has been a growing interest in exploring the connections between information theory and fundamental physics within the framework of AdS/CFT correspondence. Since Ryu and Takayanagi's conjecture \cite{Ryu:2006bv,Ryu:2006ef}, which relates the entanglement entropy of a conformal field theory to the geometry of the corresponding anti-de Sitter spacetime, the interaction between gravity and entanglement has driven significant advancements in this field. 

However, recent studies reveal that entanglement alone is insufficient to capture all features of the bulk theory \cite{Susskind:2014moa, Susskind:2014rva}. A notable example is the failure of entanglement entropy to describe the evolution of an eternal two-sided AdS black hole \cite{Maldacena:2001kr}. Specifically, when considering a thermofield double state (TFD) dual to such a black hole, the entanglement entropy of the TFD approaches equilibrium \cite{Hartman:2013qma}, while the volume of the black hole's interior continues to grow further. To address this issue, Susskind and collaborators \cite{Susskind:2014rva, Brown:2015bva, Couch:2016exn} introduced the notion of complexity, which can probe the growth of the black hole beyond the thermalization time of the entanglement entropy. Their original proposal posits that for an eternal black hole, the complexity is proportional to the spatial volume of the Einstein-Rosen bridge connecting the two boundaries.

Defining complexity within the holographic setup, however, involves subtleties that have led to various proposals, including complexity equals action (CA) \cite{Brown:2015bva}, complexity equals volume (CV) \cite{Susskind:2014rva}, complexity equals spacetime volume (CV2.0) \cite{Couch:2016exn, Omidi:2022whq}, and complexity equals anything \cite{Belin:2021bga, Belin:2022xmt, Jorstad:2023kmq, Myers:2024vve}. These conjectures aim to understand the quantum computational complexity of states in a boundary theory and their correspondence to gravitational descriptions in the bulk. The inherent subtleties are unavoidable due to similar ambiguities arising in complexity theory\footnote{For more information, see \cite{Belin:2021bga}.}.

The notion of complexity originally stems from computer science \cite{arora2009computational, moore2011nature}, where it defines the number of operations required to perform a specific task using a set of allowed simple operations. Building on this concept, various notions of complexity have emerged, such as quantum complexity \cite{Nielsen:2005mkt,nielsen2010quantum, Dowling:2006tnk, Nielsen:2006cea}, time complexity \cite{cormen2001introduction}, and holographic complexity \cite{Susskind:2014rva, Brown:2015bva, Couch:2016exn}. Generalizing the concept of complexity to different systems is a subtle task that involves making assumptions about these systems. For instance, in quantum computing, complexity is defined as the minimal number of simple unitary operations needed to transform one state into another. 

Importantly, there is no single, universal definition of complexity; rather, a family of complexity measures exists that may be multiplicatively related under certain conditions. This idea aligns with Nielsen's concept of the geometry of computations, or complexity geometry \cite{nielsen2010quantum, Dowling:2006tnk, Nielsen:2006cea}. To practically implement Nielsen's approach, various methods and techniques have been developed, such as the covariance matrix method \cite{Chapman:2018hou, Ghasemi:2021jiy, Doroudiani:2019llj, Khorasani:2023usq, Pal:2022rqq, Pal:2024tbs}, the Fubini-Study metric \cite{Chapman:2017rqy, Pal:2022ptv}, and others \cite{Bhattacharyya:2018bbv, Ali:2018fcz, Jefferson:2017sdb}.

Motivated by the significance of complexity in holography and other fields, we investigate the thermofield double (TFD) state of a quantum harmonic oscillator subjected to an external magnetic field. Specifically, we analyze the effects of the magnetic field on Nielsen complexity using the covariance matrix approach. This approach, however, is versatile and can be applied to study similar aspects in more general systems, including supersymmetric \cite{Bagchi:2001dx, Jafarov:2013cza, Fatyga:1990wx}, noncommutative \cite{Mandal:2012wp, Jing:2009lfc, BenGeloun:2009hkc, Heddar:2021sly}, relativistic \cite{Boumali:2020fqd, Falek:2017amp, Nagiyev:2023fwk, Martinez-y-Romero:1995grr}, higher derivative \cite{Mannheim:2004qz, Masterov:2015ija, Dimov:2016vvl, Pramanik:2012bh}, and other \cite{Bouguerne:2023wyl, Hamil:2022uwy, Ballesteros:2022bqx} systems.

The structure of this paper is as follows. In Section \ref{quant}, we quantize the harmonic oscillator in an external magnetic field and introduce the necessary bosonic creation and annihilation operators. In Section \ref{TFD_construct}, we construct the time-evolved thermofield double (TFD) states of the system and represent them in a suitable operator form. In Section \ref{cov_mat}, we determine the thermal covariance matrix of the system. In Section \ref{complx}, we compute the Nielsen complexity of the time-dependent TFD states and study its properties concerning temperature and the magnetic field. Furthermore, in Section \ref{lloyd_sect}, we calculate the complexity rate and demonstrate that it satisfies the Lloyd bound. Our results are summarized in Section \ref{concl}.
\section{Quantization of oscillator in an external magnetic field} \label{quant}
In this section we consider the Schr\"odinger equation for a harmonic oscillator in an external magnetic field. We find the analytic solutions and show that the system reduces to two non-interacting simple harmonic oscillators.

\subsection{The wave function}

Let's consider a spinless charged particle with mass $ m $ in a homogeneous magnetic field $ B $ oriented along the $ z $-axis, equipped by an additional harmonic potential with frequency $ \omega_0 $. Restricting the motion to the $ xy $-plane (with $ p_z = 0 $), we obtain the following Hamiltonian:
\begin{align}
H&=\frac{1}{2m} \big(\vec p -e\vec A\, \big)^2 +\frac{m \omega_0^2}{2}\big(x^2+y^2\big) \nonumber \\
&= \frac{1}{2m} \big(p_x^2+p_y^2 \big) -\frac{eB}{2m}\big(xp_y -yp_x\big) +\frac{e^2B^2}{8m} \big(x^2+y^2\big) +\frac{m \omega_0^2}{2}\big(x^2+y^2\big).
\label{Hamiltonian-1}
\end{align}
In this context, the magnetic potential $\vec{A}$ is considered in the symmetric gauge:
\begin{equation}
\vec A=\frac{B}{2} 
\left(\!\!\begin{array}{c}
-y \\
x \\
0
\end{array}\!\!\right)\!.
\end{equation}
Introducing the cyclotron frequency $\omega_c=\dfrac{eB}{m}$ and using the standard operator form of the momentum $p_i=-i \hbar \partial_i$, one has
\begin{equation}\label{Hamiltonian1}
H=-\frac{\hbar^2}{2m}(\partial_x^2 +\partial_y^2) +\frac{i\hbar\omega_c}{2} \big( x\partial_y -y\partial_x \big) +\frac{m}{8}\big(4\omega_0^2+\omega_c^2\big)  \big(x^2+y^2\big).
\end{equation}
It is useful to change to polar coordinates:
\begin{align}
x=\lambda \rho\cos\varphi, \quad y=\lambda \rho\sin\varphi,
\end{align}
where $\rho$ and $\varphi$ are dimensionless and $\lambda$ is a constant length parameter. The corresponding Laplacian and the angular momentum assume the following form:
\begin{align}
&\Delta= \partial_x^2 +\partial_y^2= \frac{1}{\lambda^2} \bigg( \partial_\rho^2 +\frac{1}{\rho}\partial_\rho +\frac{1}{\rho^2}\partial_\varphi^2 \bigg),\quad L_z=-i\hbar \big(x\partial_y -y\partial_x \big) = -i\hbar\partial_\varphi.
\end{align}
Hence, the Schr\"odinger equation, $H\Psi(\rho,\varphi)=E\Psi(\rho,\varphi)$, yields
\begin{equation}\label{SchrEq}
\bigg[-\frac{\hbar^2}{2m\lambda^2} \bigg( \partial_\rho^2 +\frac{1}{\rho}\partial_\rho +\frac{1}{\rho^2}\partial_\varphi^2 \bigg)  +\frac{i\hbar\omega_c}{2} \partial_\varphi + \frac{m}{8}\big(4\omega_0^2+\omega_c^2\big)\lambda^2 \rho^2 \bigg] \Psi= E \Psi.
\end{equation}
The Hamiltonian is independent of $\varphi$, making the angular momentum along $\varphi$ a conserved quantity, characterized by the magnetic quantum number $\ell$. Therefore, one can separate the variables using the ansatz
\begin{equation}
\Psi(\rho,\varphi)=e^{i \ell\varphi}R(\rho),
\end{equation}
which leads to the following equation for the radial part $R(\rho)$:
\begin{equation}
R''(\rho)+\frac{1}{\rho}R'(\rho) +\bigg( \frac{2m\lambda^2E}{\hbar^2}  +\frac{m\omega_c\lambda^2\ell}{\hbar} -\frac{\ell^2}{\rho^2} -\frac{m^2}{4\hbar^2}\big(4\omega_0^2+\omega_c^2\big) \lambda^4\rho^2 \bigg) R(\rho)=0.
\end{equation}
Since $\lambda$ is an arbitrary scale factor we can set its value to $\lambda=\sqrt{\dfrac{2\hbar}{m\sqrt{4\omega_0^2+\omega_c^2}}}$. After another change of variables to $\rho^2 = r$, we obtain
\begin{equation}
rR''(r) +R'(r) +\bigg( \frac{E}{\hbar\sqrt{4\omega_0^2+\omega_c^2}} +\frac{\omega_c\,\ell}{2\sqrt{4\omega_0^2+\omega_c^2}} -\frac{\ell^2}{4r} -\frac{r}{4}\bigg) R(r)=0.
\end{equation}
This is a second-order differential equation of the Laguerre type, with the solution given by
\begin{equation} 
R_{n,\ell}(r) =r^{\frac{\ell}{2}} e^{-\frac{r}{2}} L_n^{(\ell)}(r),
\end{equation}
where $L_n^{(\ell)}(r)$ are the generalized Laguerre polynomials,
\begin{equation}
L_n^{(\ell)}(r)= \frac{r^{-\ell} e^r}{n!} \frac{d^n}{dr^n} \Big( e^{-r} r^{n+\ell}\Big) = \sum\limits_{j=0}^n \frac{(-1)^j (n+\ell)!\, r^j}{j!(n-j)!(\ell+j)!}.
\end{equation}

Using the natural quantization condition: $\frac{E}{\hbar\sqrt{4\omega_0^2+\omega_c^2}} \!+\! \frac{\omega_c\,\ell}{2\sqrt{4\omega_0^2+\omega_c^2}}\!=\! n \!+\! \frac{1}{2} \!+\! \frac{\ell}{2}$, one finds the spectrum:
\begin{equation}\label{E_nl}
E_{n,\ell}= \hbar \sqrt{4\omega_0^2+\omega_c^2} \bigg(n+\frac{1}{2}\bigg) + \hbar\frac{\sqrt{4\omega_0^2+\omega_c^2} -\omega_c}{2}\, \ell,
\end{equation}
where $n$ is the principal quantum number.
Due to the normalization of Laguerre polynomials,
\begin{equation}
\int_0^\infty r^\ell e^{-r} L_n^{(\ell)}(r) L_m^{(\ell)}(r) dr= \frac{\Gamma(n+\ell+1)}{n!} \delta_{nm},
\end{equation}
and the property $n+\ell+1\geq 1$ for the gamma function, $n$ and $\ell$ take the following values:
\begin{equation}
n=0,1,2,..., \quad \ell=-n,-n+1,...,-1,0,1,2,...\,.
\end{equation}

Finally, switching back to the original $\rho$ coordinate, we can write the  complete wave function of the system:
%
\begin{align}\label{Psi_nl}
\Psi_{n,\ell}(\rho,\varphi) = \frac{1}{\lambda}\sqrt{\frac{n!}{\pi(n+\ell)!}}\, e^{i\ell \varphi} \rho^\ell e^{-\frac{\rho^2}{2}} L_n^{(\ell)} \big(\rho^2\big),
\end{align}
with natural normalization given by
\begin{equation}
\lambda^2\int_0^\infty\!\!\! \int_0^{2\pi} \Psi_{n,\ell}^*(\rho,\varphi)\Psi_{n',\ell'}(\rho,\varphi)\, \rho\, d\rho\, d\varphi= \delta_{nn'} \delta_{\ell \ell'}.
\end{equation}
\subsection{Fock space}
In order to construct the Fock space it is convenient to introduce the following creation and annihilation operators:
\begin{align} \label{defAB}
&a=-\frac{e^{i\varphi}}{2} \bigg(\rho +\partial_\rho +\frac{i}{\rho} \partial_\varphi \bigg), \quad  a^\dagger=-\frac{e^{-i\varphi}}{2} \bigg(\rho -\partial_\rho +\frac{i}{\rho} \partial_\varphi \bigg) \nonumber \\
&b=\frac{e^{-i\varphi}}{2} \bigg(\rho + \partial_\rho -\frac{i}{\rho} \partial_\varphi \bigg), \quad \,\, b^\dagger=\frac{e^{i\varphi}}{2} \bigg(\rho -\partial_\rho -\frac{i}{\rho} \partial_\varphi \bigg), 
\end{align}
which satisfy the standard commutation relations:
\begin{equation} \label{commAB}
 [a,a^\dagger]=1 =[b,b^\dagger], \quad [a,b]=0.
\end{equation}
The action of these operators on the wave function (\ref{Psi_nl}) is given by:
\begin{align}
&a^\dagger \Psi_{n,\ell} = \sqrt{n+1}\, \Psi_{n+1,\ell-1}, \quad  b^\dagger \Psi_{n,\ell} = \sqrt{n+\ell+1}\, \Psi_{n,\ell+1}, \nonumber\\ 
&a \Psi_{n,\ell}= \sqrt{n}\, \Psi_{n-1,\ell+1}, \,\quad\qquad b \Psi_{n,\ell}= \sqrt{n+\ell\,}\, \Psi_{n,\ell-1} .
\end{align}
The action of $b^\dagger$ increases the number $\ell$ by one unit, while preserving $n$. The operator $a^\dagger$ simultaneously increases $n$ and decreases $\ell$ by one unit. We can simplify the notations by introducing a new shifted quantum number $k = n+\ell$. The relation between the existing wave function $\Psi_{n,\ell}$ and the new wave function $\Phi_{n,k}$ is given by:
\begin{equation}
\Psi_{n, k-n}= \Phi_{n,k} = \frac {(a^\dagger )^n (b^\dagger)^{k} } {\sqrt {n!k!} } \Phi_{0,0}, \quad \Psi_{0,0}=\Phi_{0,0} = \frac{e^{-\frac{\rho^2}{2} }}{\lambda\sqrt{\pi}}. 
\end{equation}
Now the operators $a$ and $a^\dagger$ affect only $n$, and the operators $b$ and $b^\dagger$ change only $k$:
\begin{align}
&a^\dagger \Phi_{n,k} = \sqrt{n+1}\, \Phi_{n+1,k}, \quad  b^\dagger \Phi_{n,k} = \sqrt{k+1}\, \Phi_{n,k+1}, \nonumber\\ 
&a \Phi_{n,k}= \sqrt{n}\, \Phi_{n-1,k}, \quad\qquad b \Phi_{n,k}= \sqrt{k}\, \Phi_{n,k-1} .
\end{align}
The energy in this case can be written in terms of the $n, k$ indices in the following way:
\begin{equation} \label{energy}
E_{n,k}=\hbar\omega_1\bigg(n +\frac{1}{2} \bigg) +\hbar\omega_2\bigg(k +\frac{1}{2} \bigg), \quad a^\dagger a\, \Phi_{n,k} = n \Phi_{n,k}, \quad b^\dagger b\, \Phi_{n,k} = k \Phi_{n,k}.
\end{equation}
These are the eigenvalues of a Hamiltonian for two non-interacting harmonic oscillators with frequencies: 
\begin{equation}\label{frequencies}
\omega_1=\frac{\sqrt{4\omega_0^2 +\omega_c^2} +\omega_c}{2}, \quad \omega_2=\frac{\sqrt{4\omega_0^2 +\omega_c^2} -\omega_c}{2}.
\end{equation}
Note that $\omega_1 \geq \omega_2$, with equality occurring only when the magnetic field is turned off $(\omega_1=\omega_2=\omega_0, \,\omega_c=0)$.

\section{Construction of TFD state} \label{TFD_construct}
To construct the TFD state we apply the standard approach. We double the Hilbert space to left/right sectors, which commute. In the $\Phi_{n,k}\equiv|n,k \rangle$ basis we can write the TFD state as:
\begin{align}\label{TFD}
\nonumber |T\!F\!D \rangle&= \frac{1}{\sqrt{Z}}\sum\limits_{n=0}^\infty \sum\limits_{k=0}^\infty e^{-\frac{\beta}{2}E_{n,k}} |n,k \rangle_L |n,k \rangle_R \\ 
&=\frac{e^{-\frac{\beta\hbar }{4}(\omega_1+\omega_2)} }{\sqrt{Z}} \sum\limits_{n=0}^\infty \sum\limits_{k=0}^\infty e^{-\frac{\beta\hbar\omega_1 }{2}n} e^{-\frac{\beta\hbar\omega_2 }{2}k} \frac{\big(a^\dag_L a^\dag_R\big)^n}{n!} \frac{\big(b^\dag_L b^\dag_R\big)^k}{k!}  |0,0 \rangle_L |0,0 \rangle_R \nonumber \\
&=\frac{e^{-\frac{\beta\hbar }{4}(\omega_1+\omega_2)} }{\sqrt{Z}} \exp \!\big(  e^{-\frac{\beta\hbar\omega_1 }{2}} a^\dag_L a^\dag_R + e^{-\frac{\beta\hbar\omega_2 }{2}}b^\dag_L b^\dag_R \big) |0,0 \rangle_L |0,0 \rangle_R,
\end{align}
where the energy $E_{n,k}$ is defined in Eq. \eqref{energy}. The operators\footnote{The left and the right operators commute: $[a_L,a_R^\dagger]=0$, $[b_L,b_R^\dagger]=0$.} $a_{L/R}$ and $b_{L/R}$ are defined in \eqref{defAB} and the partition function $Z$ follows from the normalization condition $\langle T\!F\!D|T\!F\!D \rangle=1$:
\begin{equation}\label{PartF}
Z=\frac{e^{-\frac{\beta\hbar}{2}(\omega_1+\omega_2)}}{ \big(1-e^{-\beta\hbar\omega_1}\big) \big(1-e^{-\beta\hbar\omega_2}\big) }= \frac{1}{4 \sinh \frac{\beta\hbar \omega_1}{2} \sinh \frac{\beta \hbar \omega_2}{2}}.
\end{equation}
This is e exactly the partition function of two decoupled harmonic oscillators. Expressing the TFD as a unitary operator acting on the vacuum, we find\footnote{See the Appendix of \cite{chapman2019complexity}.}
\begin{equation}\label{TFD_ti}
|T\!F\!D \rangle =\exp\big[\alpha_1\big(a_L^\dagger a_R^\dagger -a_L a_R \big) +\alpha_2\big(b_L^\dagger b_R^\dagger -b_L b_R\big)\big] |0,0 \rangle_L |0,0 \rangle_R,
\end{equation}
where the coefficients $\alpha_{i}$, $i=1,2$, are written by \cite{chapman2019complexity}:
\begin{equation}\label{eqAlphai}
\tanh\alpha_{i}=e^{-\frac{\beta\hbar\omega_{i}}{2}}.
\end{equation}
Since the TFD state \eqref{TFD_ti} includes only $\alpha_1$ and $\alpha_2$ as parameters we can simplify notations by $|T\!F\!D\rangle =|\alpha_1,\alpha_2\rangle$.

We can extend (\ref{TFD}) and (\ref{TFD_ti}) to the time-dependent case in the following way:
\begin{align}\label{TFD(t)}
&|\alpha_1,\alpha_2,t \rangle= \frac{1}{\sqrt{Z}}\sum\limits_{n=0}^\infty \sum\limits_{k=0}^\infty e^{-\frac{\beta}{2}E_{n,k}}\, e^{-\frac{i}{\hbar} E_{n,k}t} \, |n,k \rangle_L |n,k \rangle_R \nonumber \\
&=\exp\big[\alpha_1\big( e^{-i\omega_1 t}a_L^\dagger a_R^\dagger -e^{i\omega_1 t}a_L a_R \big) +\alpha_2\big( e^{-i\omega_2 t}b_L^\dagger b_R^\dagger -e^{i\omega_2 t}b_L b_R\big)\big] |0,0 \rangle_L |0,0 \rangle_R  ,
\end{align}
where $\alpha_i$ are the same as in (\ref{eqAlphai}).
One can introduce standard positions and momenta corresponding to the creation/annihilation operators by:
\begin{align}
a_{L/R}=\sqrt{\frac{m\omega_1}{2\hbar}} \bigg( X_{1\,L/R} +i\frac{P_{1\,L/R}}{m\omega_1} \bigg), \quad
b_{L/R}=\sqrt{\frac{m\omega_2}{2\hbar}} \bigg( X_{2\,L/R} +i\frac{P_{2\,L/R}}{m\omega_2} \bigg).
\end{align}
A subsequent transformation to a new $\pm$ basis yields:
\begin{align}\label{lightcone}
&X_{1\pm} =\frac{1}{\sqrt{2}} \big(X_{1L} \pm X_{1R} \big), \quad P_{1\pm} =\frac{1}{\sqrt{2}} \big(P_{1L} \pm P_{1R} \big), \nonumber \\
&X_{2\pm} =\frac{1}{\sqrt{2}} \big(X_{2L} \pm X_{2R} \big), \quad P_{2\pm} =\frac{1}{\sqrt{2}} \big(P_{2L} \pm P_{2R} \big).  
\end{align} 
Hence, the argument of the unitary operator from \eqref{TFD_ti} takes the form:
\begin{equation}
\alpha_1\big(a_L^\dagger a_R^\dagger -a_L a_R \big) +\alpha_2\big(b_L^\dagger b_R^\dagger -b_L b_R\big)= -i\alpha_1  K_{1+} +i\alpha_1 K_{1-} -i\alpha_2 K_{2+} +i\alpha_2  K_{2-},
\end{equation}
where the $ K_{i\,\pm}$ operators are:
\begin{equation}\label{Ki}
	K_{i\pm}=  \frac{1}{2\hbar} \big( X_{i\pm}P_{i\pm} +P_{i\pm}X_{i\pm} \big).
\end{equation}
This leads to the following form of the time-independent TFD:
\begin{equation}\label{TFD_titi}
|\alpha_1,\alpha_2 \rangle =e^{-i\alpha_1 K_{1+}} |0_{1+}\rangle \otimes e^{i\alpha_1 K_{1-}} |0_{1-}\rangle \otimes e^{-i\alpha_2 K_{2+}} |0_{2+}\rangle \otimes e^{i\alpha_2 K_{2-}} |0_{2-}\rangle.
\end{equation}
Therefore, in this basis, the unitary operator \eqref{TFD_ti} completely decouples into four separate parts, since all $K_{i\, \pm}$ commute with each other. 

Similarly, we can calculate the new time-dependent unitary operator from \eqref{TFD(t)} as:
\begin{align}\label{K_t}
\alpha_1\big( e^{-i\omega_1 t}a_L^\dagger a_R^\dagger -e^{i\omega_1 t}a_L a_R \big) +\alpha_2\big( e^{-i\omega_2 t}b_L^\dagger b_R^\dagger -e^{i\omega_2 t}b_L b_R\big) =\nonumber \\ 
=-i\alpha_1 {K}_{1+}(t) +i\alpha_1 {K}_{1-}(t)-i\alpha_2{K}_{2+}(t) +i\alpha_2{K}_{2-}(t),
\end{align}
where the time-dependent generalizations of the operators $K_i$ \eqref{Ki} can be expressed as:
\begin{equation} \label{Ki(t)}
\hat K_{i\pm}(t) =\frac{1}{2\hbar}\cos\omega_it \big( X_{i\pm}P_{i\pm} +P_{i\pm}X_{i\pm} \big) +\frac{1}{2\hbar} \sin\omega_it \bigg( m\omega_i X_{i\pm}^2 -\frac{P_{i\pm}^2}{m\omega_i} \bigg).
\end{equation}
At $t=0$ one has $ K_{i\pm}(0)= K_{i\pm}$ (see Eq. \eqref{Ki}).
The time-dependent TFD state  \eqref{TFD(t)} now becomes:
\begin{equation}\label{TFDtd}
|\alpha_1,\alpha_2,t \rangle =e^{-i\alpha_1 K_{1+}(t)} |0_{1+}\rangle \otimes e^{i\alpha_1 K_{1-}(t)} |0_{1-}\rangle \otimes e^{-i\alpha_2 K_{2+}(t)} |0_{2+}\rangle \otimes e^{i\alpha_2 K_{2-}(t)} |0_{2-}\rangle,
\end{equation}
which also decouples into four disjoint parts.

\section{Covariance matrix} \label{cov_mat}

In this section, we use the covariance matrix approach for computing Nielsen complexity, as introduced in \cite{chapman2019complexity}. This method is particularly useful for estimating complexity when both the initial and final states of the evolution are Gaussian, which is the case for the TFD state. The approach can be summarized in a few simple steps. First, represent the TFD state as generated by some unitary operator acting on the vacuum, $\ket{T\!F\!D}\! =\! U(t) \ket{0}$. Next, consider a set of generators $\xi^r$ that span all generators of the problem. Based on these generators, construct the covariance matrix
\begin{equation}
    G^{rs} = \bra{\psi} \xi^r \xi^s + \xi^s \xi^r \ket{\psi},
\end{equation}
where $\psi$ is an arbitrary state. Using this definition one can define the relative covariance matrix between two states by
\begin{equation}\label{GTGR}
    \Delta  = G_T. G^{-1}_R, 
\end{equation}
where $G^{-1}_R$ is the inverse of the covariance matrix of the reference state, while $G_T$ is the covariance matrix of the target state. Finally, one can express the Nielsen complexity using the relative covariance matrix $\Delta$ via the Frobenius norm

\begin{equation}
    \mathcal{C}(t) = \frac{1}{2} \sqrt{\mathrm{Tr} (\ln \Delta)^2} .
\end{equation}
Thus, the problem of finding complexity reduces to determining the eigenvalues of $\Delta$ (for details see \cite{chapman2019complexity} and references therein).

\subsection{Covariance matrix for the time-independent TFD state}

The covariance matrix for this problem can be expressed in terms of the set of operators
\begin{equation}
\vec\xi =\vec\xi_{1+} \oplus\, \vec\xi_{1-} \oplus\, \vec\xi_{2+} \oplus\, \vec\xi_{2-} =(X_{1+},P_{1+},X_{1-},P_{1-},X_{2+},P_{2+},X_{2-},P_{2-})^T\equiv (\xi^r), 
\end{equation}
where the two component vectors $\xi_{i\pm}$ are
\begin{align}
\vec\xi_{i\pm}=(X_{i\pm},P_{i\pm})^T.
\end{align}
The vacuum covariance matrix $O_0^{rs}, (r,s=1,...,8)$ is written by
\begin{equation}
O_0^{rs} =\langle 0|\xi^r \xi^s |0\rangle = \frac{1}{2}\langle 0|\big\{ \xi^r, \xi^s \big\} |0 \rangle +\frac{1}{2} \langle 0| \big[\xi^r, \xi^s\big] |0 \rangle = \frac{\hbar}{2}\big( G^{rs}_0 +i\Omega^{rs}_0 \big),
\end{equation}
where one has $8\times 8$ block-diagonal matrices:
\begin{equation}\label{vacCovM}
	G_0= \left(\!\!
	\begin{array}{cccc}
		G_{01} & 0 & 0 & 0 \\
		0 & G_{01} & 0 & 0 \\
		0 & 0 & G_{02} & 0 \\
		0 & 0 & 0 & G_{02} \\
	\end{array}
	\!\!\right)\!, \quad  \Omega_0 = \left(\!\!
	\begin{array}{cccc}
		\Omega_{01} & 0 & 0 & 0 \\
		0 & \Omega_{01} & 0 & 0 \\
		0 & 0 & \Omega_{02} & 0 \\
		0 & 0 & 0 & \Omega_{02} \\
	\end{array}
	\!\!\right),
\end{equation}
with $2\times 2$ submatrices given by
\begin{equation}
	G_{0i}= \left(\!\!
	\begin{array}{cc}
		\frac{1}{m\omega_i}  & 0 \\
		0 & m\omega_i \\
	\end{array}
	\!\!\right)\!, \quad
	\Omega_{0i} = \left(\!\!
	\begin{array}{cc}
		0 & 1 \\
		-1 & 0  \\
	\end{array}
	\!\!\right)\!.
\end{equation}

The action of the unitary operators $ U_{i\pm} =e^{\mp i \alpha_i  K_{i\pm}}$ on the vacuum generate the TFD state  \eqref{TFD_titi}:
\begin{equation}
|\alpha_1,\alpha_2\rangle =  U_{1+} |0_{1+}\rangle \otimes  U_{1-} |0_{1-}\rangle \otimes  U_{2+} |0_{2+}\rangle\otimes  U_{2-}|0_{2-}\rangle . 
\end{equation}
One can find the matrix representations $\mathcal U_{i\pm}$ and $\mathcal K_{i\pm}$ of each of the operators by the relations (there is no sum over $i$):
\begin{equation}
	 U_{i\pm}^\dagger \xi_{i\pm}^a  U_{i\pm} =\mathcal U^{\,\,\,\,a}_{i\pm,b\,} \xi_{i\pm}^b,\quad [i K_{i\pm},\xi_{i\pm}^a] =\big(\mathcal K_{i\pm}. \vec\xi_{i\pm} \big)^a =\mathcal K^{\,\,\,\,\,a}_{i\pm,b\,} \xi_{i\pm}^b, \quad a,b=1,2.
\end{equation}
Now we calculate the following product\footnote{There is no sum over $i$.}:
\begin{align}
U_{i\pm}^\dagger \xi_{i\pm}^a  U_{i\pm}&= e^{\pm i\alpha_i K_{i\pm}} \xi_{i\pm}^a e^{\mp i\alpha_i K_{i\pm}} =\sum_{n=0}^\infty \frac{(\pm\alpha_i)^n}{n!} [i K_{i\pm},\xi_{i\pm}^a]_{(n)} \nonumber \\ 
&=\sum_{n=0}^\infty \frac{(\pm\alpha_i)^n}{n!}  \big(\mathcal K_{i\pm}^n .\vec\xi_{i\pm} \big)^a  =\Big( e^{\pm\alpha_i\mathcal K_{i\pm}} .\vec\xi_{i\pm} \Big)^{\!a} =\Big( \mathcal U_{i\pm} .\vec\xi_{i\pm} \Big)^{\!a} =\mathcal U^{\,\,\,\,a}_{i\pm,b\,} \xi_{i\pm}^b,
\end{align}
where  $\big[ i\hat K_+,\xi^a_{1+} \big]_{(n)}$ denotes the $n$-th nested commutator.
We find the commutators between $ K_{i\pm}$ and $\xi_{i\pm}^a$:
\begin{equation}
	[i K_{i\pm},X_{i\pm}]=   X_{i\pm}, \quad  [i K_{i\pm},P_{i\pm}]= - P_{i\pm}, 
\end{equation}
which set the form of the matrices $\mathcal K_{1+}=\mathcal K_{1-}=\mathcal K_{2+}=\mathcal K_{2-}=\mathcal K$:
\begin{equation}
	\mathcal K =\left(\!
	\begin{array}{cc}
		1 \!& 0  \\
		0 \!& -1 
	\end{array}
	\!\right)\!.
\end{equation}
After exponentiating the above matrix we find $\mathcal U_{i\pm}$:
\begin{equation}
\mathcal U_{i\pm} =e^{\pm\alpha_i \mathcal K}= \mathbb{1} \cosh \alpha_i \pm \mathcal K \sinh \alpha_i =\left(\!\!
\begin{array}{cc}
		e^{\pm \alpha_i} \!& 0 \\
		0 \!& e^{\mp\alpha_i} \\
\end{array}
\!\!\!\right)\!.	
\end{equation}

The components of the full TFD covariance matrix follow from $O^{rs}=\langle \alpha_1,\alpha_2|\xi^r \xi^s |\alpha_1,\alpha_2\rangle$. 
Here we can show how to calculate the upper-left block: 
\begin{align}
\nonumber O^{ab}_{1+}&=\langle \alpha_1,\alpha_2|\xi^a_{1+} \xi^b_{1+} |\alpha_1,\alpha_2\rangle = \langle 0_{1+}|  U^\dagger_{1+} \xi^a_{1+} \xi^b_{1+}  U_{1+} |0_{1+}\rangle 
\\
\nonumber
&= \mathcal U^{\,\,\,\,\,a}_{1+,c\,} \langle 0_{1+}| \xi^c_{1+} \xi^d_{1+}  |0_{1+}\rangle\,
\mathcal U^{\,\,\,\,\,b}_{1+,d} = \mathcal U^{\,\,\,\,\,a}_{1+,c\,} O_{01}^{cd}\, \mathcal U^{\,\,\,\,\,b}_{1+,d}  \nonumber \\
&=\frac{\hbar}{2} \Big( \mathcal U^{\,\,\,\,\,a}_{1+,c\,} G_{01}^{cd}\, \mathcal U^{\,\,\,\,\,b}_{1+,d}  +i\,\mathcal U^{\,\,\,\,\,a}_{1+,c\,} \Omega_{01}^{cd}\, \mathcal U^{\,\,\,\,\,b}_{1+,d}   \Big) =\frac{\hbar}{2} \Big( G_{1+}^{ab}  +i \Omega_{1+}^{ab}  \Big). 
\end{align}
Similar calculations work  for each of the subblocks of matrices $O_{1-},O_{2+}$ and $O_{2-}$. Therefore, we can write the explicit form of $G_{i\pm}$ and $\Omega_{i\pm}$ matrices  as:
\begin{equation}
G_{i\pm}=\mathcal U_{i\pm}.G_{0i}.\,\mathcal U_{i\pm}^T= \left(\!
\begin{array}{cc}
	\dfrac{e^{\pm2\alpha_i}}{m\omega_i} \!& 0 \\
	0 \!& m\omega_i e^{\mp2\alpha_i}
\end{array}
\!\!\right)\!,\quad
\Omega_{i\pm}=\mathcal U_{i\pm}.\Omega_{0i}.\,\mathcal U_{i\pm}^T =\Omega_{0i}. 
\end{equation}
Putting everything together, we find the full time-independent TFD covariance matrix:
\begin{equation}
G= \left(\!
\begin{array}{cccc}
G_{1+}& 0 & 0 & 0  \\
0 & G_{1-} & 0 & 0  \\
0 & 0 & G_{2+} & 0  \\
0 & 0 & 0 & G_{2-}  \\
\end{array}
\!\!\right)\!. 
\end{equation}

This concludes the construction of the time-independent TFD case. 

\subsection{Covariance matrix for the time-dependent TFD state}

The analysis of the time-dependent case follows the same steps.
We can create the time-dependent TDF state (\ref{TFDtd}) by acting on the vacuum with the unitary operators $ U_{i\pm}(t)\!=\!e^{\mp i \alpha_i K_{i\pm}(t)}$:
\begin{equation}
|\alpha_1,\alpha_2,t\rangle =  U_{1+}(t) |0_{1+}\rangle \otimes  U_{1-}(t) |0_{1-}\rangle \otimes U_{2+}(t) |0_{2+}\rangle\otimes  U_{2-}(t)|0_{2-}\rangle,
\end{equation}
where the operators $ K_{i\pm}(t)$ are defined in Eq. \eqref{Ki(t)}. Here we introduce the matrices $\mathcal U_{i\pm}(t)$ and  $\mathcal K_{i\pm}(t)$ by the following relations:
\begin{equation}
 U_{i\pm}^\dagger(t)\, \xi_{i\pm}^a  U_{i\pm}(t) =\mathcal U^{\,\,\,\,a}_{i\pm,b}(t)\, \xi_{i\pm}^b, \quad [i K_{i\pm}(t),\xi_{i\pm}^a] =\mathcal K^{\,\,\,\,\,a}_{i\pm,b}(t)\, \xi_{i\pm}^b.
\end{equation}
The commutation relations between $ K_{i\pm}(t)$ and $\xi_{i\pm}^a$ yield:
\begin{align}
[i K_i(t),X_{i\pm}] &= 
 \cos\omega_it\, X_{i\pm} -\frac{\sin\omega_it}{m \omega_i} P_{i\pm}, \nonumber \\
[i K_i(t),P_{i\pm}] &= -\cos\omega_it\, P_{i\pm} -m\omega_i \sin\omega_it\, X_{i\pm} ,
\end{align}
hence one can derive the explicit form of the matrices $\mathcal K_{i+}(t) =\mathcal K_{i-}(t)= \mathcal K_{i}(t)$:
\begin{equation}
 \mathcal K_i(t) =\left(\!\!
\begin{array}{cc}
\cos\omega_it \!& -\dfrac{\sin\omega_it}{m\omega_i} \\
-m\omega_i \sin\omega_it \!& -\cos\omega_it \\
\end{array}
\!\right)\!.
\end{equation}
The corresponding matrices  $\mathcal U_{i\pm}(t)$ follow directly,
\begin{align}
\mathcal U_{i\pm}(t) &=e^{\pm\alpha_i \mathcal K_i(t)}= \mathbb{1} \cosh \alpha_i \pm \mathcal K_i(t) \sinh \alpha_i  \nonumber \\
&=\left(\!\!
\begin{array}{cc}
\cosh\alpha_i \pm \sinh \alpha_i \cos\omega_it & \mp \dfrac{ \sinh \alpha_i \sin\omega_it}{m\omega_i}  \\
\mp m\omega_i\, \sinh\alpha_i \sin\omega_it & \cosh\alpha_i \mp \sinh\alpha_i \cos\omega_it \\
\end{array}
\!\!\right)\!.	
\end{align}
The components of the time-dependent covariance matrix are naturaly defined by $O^{rs}(t)=\langle \alpha_1,\alpha_2,t|\xi^r \xi^s |\alpha_1,\alpha_2,t\rangle$, which leads to the following expressions for $\Omega_{i\pm}(t)= \Omega_{0i}$ and $G_{i\pm}(t)$:
%
%
\begin{align}
G_{i\pm}(t) \!=\mathcal U_{i\pm}(t).G_{0i}.\,\mathcal U_{i\pm}^T(t) \!= \!\!\left(\!\!\!
\begin{array}{cc}
\dfrac{ \cosh\! 2\alpha_i \pm \sinh\! 2\alpha_i \cos\omega_i t}{m\omega_i}  \!\!\!&\!\!\! \mp\, \sinh 2\alpha_i \sin\omega_i t \\
\mp\, \sinh\! 2\alpha_i \sin\omega_i t \!\!\!&\!\!\! m\omega_i \big(\! \cosh\! 2\alpha_i \mp \sinh\! 2\alpha_i \cos\omega_i t \big) \\ \end{array}
\!\!\!\!\right)\!\!.
\end{align}
Therefore, the full time-dependent covariance matrix assumes to the following form
\begin{equation}\label{GT(t)}
G(t)= \left(\!\!
	\begin{array}{cccc}
		G_{1+}(t) & 0 & 0 & 0 \\
		0 & G_{1-}(t) & 0 & 0 \\
		0 & 0 & G_{2+}(t) & 0 \\
		0 & 0 & 0 & G_{2-}(t) \\
	\end{array}
	\!\!\right)\!.
\end{equation}

This concludes the construction of the time-evolved TFD for our problem. The final step is to compute the complexity of the system.

\section{Complexity}
\label{complx}
In this section, we proceed with our analysis by calculating the relative covariance matrix and determining its eigenvalues. Next, we employ the Frobenius norm of a particular generator—constructed from the eigenvalues of the relative covariance matrix—to determine the complexity of the harmonic oscillator, as outlined in \cite{chapman2019complexity}. Finally, we explore how complexity behaves across different temperature and magnetic field regimes.

\subsection{Relative covariance matrix and complexity}

We choose the target state $G_T$ form \eqref{GTGR} to be the full $8\times 8$ time-dependent covariance $G(t)$, defined in \eqref{GT(t)}. The reference state $G_{R}$ is given by the vacuum covariance matrix \eqref{vacCovM} with an  initial frequency $\omega_{Ri}\neq \omega_i$:
\begin{equation}
G_R=\left(\!\!
\begin{array}{cccc}
	G_{R1} & 0 & 0 & 0 \\
	0 & G_{R1} & 0 & 0 \\
	0 & 0 & G_{R2} & 0 \\
	0 & 0 & 0 & G_{R2} \\
\end{array}
\!\!\right), \quad   G_{Ri}=\left(\!\!\!
\begin{array}{cc}
	\frac{1}{m\omega_{Ri}} & 0\\
	0 & m\omega_{Ri}
\end{array}
\!\!\right)\!.
\end{equation}

The relative covariance matrix $\Delta$ has the form:
\begin{equation}\label{eqRelCovMatr}
\Delta(t)=G(t). G_R^{-1}=\left(\!\!
\begin{array}{cccc}
\Delta_{1+}(t) & 0 & 0 & 0 \\
0 & \Delta_{1-}(t) & 0 & 0 \\
0 & 0 & \Delta_{2+}(t) & 0 \\
0 & 0 & 0 & \Delta_{2-}(t) \\
\end{array}
\!\!\right)\!,
\end{equation}
where the sub-matrices $\Delta_{i\pm}(t)$ are written by
\begin{equation}
\Delta_{i\pm}(t)=G_{i\pm}(t).G^{-1}_{Ri}= \!\!\left(\!\!\!
\begin{array}{cc}
	\dfrac{\omega_{Ri}}{\omega_i} \big( \cosh\! 2\alpha_i \pm \sinh\! 2\alpha_i \cos\omega_i t \big) \!\!&\!\! \mp\, \dfrac{ \sinh 2\alpha_i \sin\omega_i t}{m\omega_{Ri}} \\
	\mp\, m\omega_{Ri}\, \sinh 2\alpha_i \sin\omega_i t \!\!&\!\! \dfrac{\omega_i}{\omega_{Ri}} \big(\! \cosh\! 2\alpha_i \mp \sinh\! 2\alpha_i \cos \omega_i t \big) \\ \end{array}
\!\!\!\!\right)\!\!.	
\end{equation}
The relative covariance matrix has 8 positive eigenvalues of the following form:
\begin{align}\label{eigenvalues}
&e_1=A_{1+}-\sqrt{A_{1+}^2-1}\,, \quad e_2=A_{1+}+\sqrt{A_{1+}^2-1}\,, \nonumber \\
&e_3=A_{1-}-\sqrt{A_{1-}^2-1}\,, \quad e_4=A_{1-}+\sqrt{A_{1-}^2-1}\,, \nonumber \\
&e_5=A_{2+}-\sqrt{A_{2+}^2-1}\,, \quad e_6=A_{2+}+\sqrt{A_{2+}^2-1}\,, \nonumber \\
&e_7=A_{2-}-\sqrt{A_{2-}^2-1}\,, \quad e_8=A_{2-}+\sqrt{A_{2-}^2-1}\,,
\end{align}
where the functions $A_{i \pm}$ carry the time dependence:
\begin{equation}\label{AAA}
A_{i\pm}=\frac{1}{2\omega_{Ri}\omega_i} \Big( \big(\omega_{Ri}^2+\omega_i^2\big) \cosh 2\alpha_i \pm \big(\omega_{Ri}^2-\omega_i^2\big) \sinh 2\alpha_i \cos \omega_i t \Big).
\end{equation}
One notes that if $\omega_{Ri}=\omega_i$ the eigenvalues are time-independent. This situation also occurs at very low temperature $(\alpha_i\to 0)$. We can write the relative covariance matrix in a diagonal form by $\Delta={\rm{diag}} ( e_1,...,e_8 )$. The matrix $\sqrt{\Delta}$, which is a linear map between the states, 
\begin{equation}
 \big( \sqrt{\Delta} \big)^2.G_R= \Delta.G_R= G(t).G_R^{-1}.G_R=G(t),
\end{equation}
can be expressed in an exponential form $\sqrt{\Delta}=e^M$, where the generator $M$ is given by $M=\ln \sqrt{\Delta} =\frac{1}{2}\ln \Delta$ $=\frac{1}{2}\, {\rm{diag}} \big( \ln e_1, ...,\ln e_8 \big)$. The geodesic distance between the reference state $G_R$ and the target state $G(t)$ is equal to the complexity. Following \cite{chapman2019complexity} we can find it using the Frobenius norm of the matrix $\mathcal C(t)=|\!|M|\!|=  \sqrt{{\rm{Tr}} M^2}$, i.e.
\begin{align}\label{Complexity}
\mathcal C(t) &= \frac{1}{2} \sqrt{ {\rm{Tr}} (\ln \Delta)^2 } =\frac{1}{2} \sqrt{ \sum_{s=1}^8 (\ln e_s)^2\, } .
\end{align}
Due to Eq. \eqref{AAA} complexity $\mathcal C(t)$ is a periodic function in time. In general, its period has a nontrivial dependence on the frequencies $\omega_{1,2}$. However, in the weak/strong magnetic field regimes it is possible to derive an approximate formulae for the period of the complexity oscillations, as shown in Subsection \ref{eqMFA}.

\subsection{Complexity for $\omega_{Ri}=\omega_i$}

If we chose the reference frequencies to be equal to the target ones $\omega_{Ri}=\omega_i$, the complexity \eqref{Complexity} does not depend on time and it simplifies to
\begin{equation}\label{C_wR=w}
\mathcal C=2 \sqrt{\alpha_1^2 +\alpha_2^2\, } =2\sqrt{{\rm{arcth}}^2 e^{-\frac{\beta\hbar\omega_1}{2}} + {\rm{arcth}}^2  e^{-\frac{\beta\hbar\omega_2}{2}}  }.
\end{equation}
Considering complexity as a function of the temperature we discern two regimes.
For low temperatures, $\beta \hbar\omega_i =\dfrac{\hbar\omega_i}{kT}\gg 1$, the asymptotic expansion of \eqref{C_wR=w} yields
\begin{equation}
\mathcal C\approx 2 \sqrt{ e^{-\beta\hbar\omega_1} +e^{-\beta\hbar\omega_2} }, \quad \lim_{\beta\to\infty} \mathcal C=0.
\end{equation}
In this situation the complexity is exponentially suppressed by the temperature and goes to zero for $\beta\to \infty$. 
For high temperature, $\beta \hbar\omega_i =\dfrac{\hbar\omega_i}{kT}\ll 1$, complexity takes the asymptotic form
\begin{equation}
\mathcal C\approx \sqrt{ \ln^2 \frac{4}{\beta\hbar\omega_1} +\ln^2 \frac{4}{\beta\hbar\omega_2} }, \quad \lim_{\beta\to 0} \mathcal C= \infty.
\end{equation}
Here one has a logarithmic divergence for high temperature. 

Turning off the external magnetic field $\omega_c=0$, leads to $\omega_1=\omega_2= \omega_0$, hence:
\begin{equation}
\mathcal C=2\sqrt{2}\,{\rm{arcth}} \big(e^{-\frac{\beta\hbar\omega_0}{2}}\big) .
\end{equation}
The above formula represents the complexity of two harmonic oscillators with equal frequencies.

\subsection{Complexity for $\omega_{Ri}\neq\omega_i$}

\subsubsection{Temperature analysis}

For very low temperature $\beta\to \infty$, the eigenvalues \eqref{eigenvalues} become time-independent, since $\alpha_i\to 0$. Therefore, complexity \eqref{Complexity} is only frequency dependent quantity, which saturates to a minimum positive value (the red line on Figs.\,\ref{w1>>w2}, \ref{w1>>w2c} and \ref{w1_w2}):
\begin{equation}\label{red}
\lim_{\beta\to \infty}\mathcal C(t)=\sqrt{ \ln^2 \frac{\omega_{R1}}{\omega_1} +\ln^2\frac{\omega_{R2}}{\omega_2} }\,.
\end{equation}
For high temperature $\beta\hbar\omega_i\ll 1$ the asymptotic behaviour of complexity is determined by
\begin{equation} \label{C_Tinf}
{\cal C}(t)\, \approx \,\sqrt{ \frac{1}{2}\,\sum\limits_{i = 1}^2  \,\big( C_{i+} + C_{i-} \big)} ,\quad C_{i\pm} = \ln ^2 \frac{2\big[ (\omega_{Ri}^2 + \omega_i^2) \pm (\omega_{Ri}^2 -\omega_i^2) \cos\omega_it \big]} {\beta \hbar \omega_{Ri\,} \omega_i^2}.
\end{equation}
As to be expected $\mathcal C(t)$ is unbounded from above: $\mathcal C(t) \to \infty$ for $\beta \to 0$.

\subsubsection{Magnetic field analysis}\label{eqMFA}

We investigate the two sectors of strong/weak magnetic field.  It is useful to set $\hbar = 1$ and $\omega_{R1} = \omega_{R2} = 1$.
\\

\begin{figure}[H]
\centering \hspace{-1.0cm}
\begin{subfigure}{0.4\textwidth}
\includegraphics[width=8.3cm,height=5.55cm]{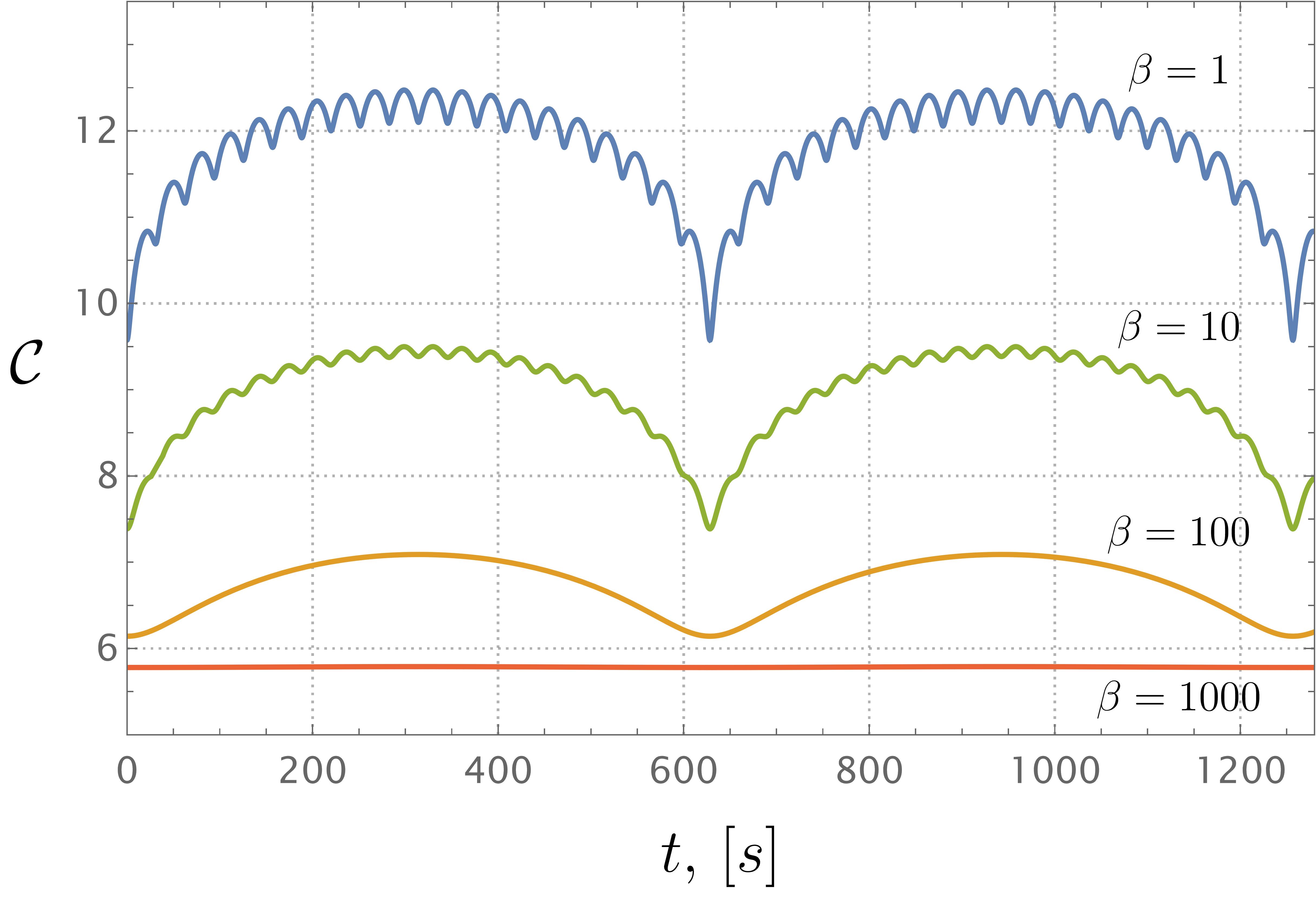}
\caption{Complexity $\mathcal C(t)$ at strong $B$.}\label{w1>>w2}
\end{subfigure}
\hspace{1.5 cm}
\begin{subfigure}{0.4\textwidth}
\includegraphics[width=8.3cm,height=5.5cm]{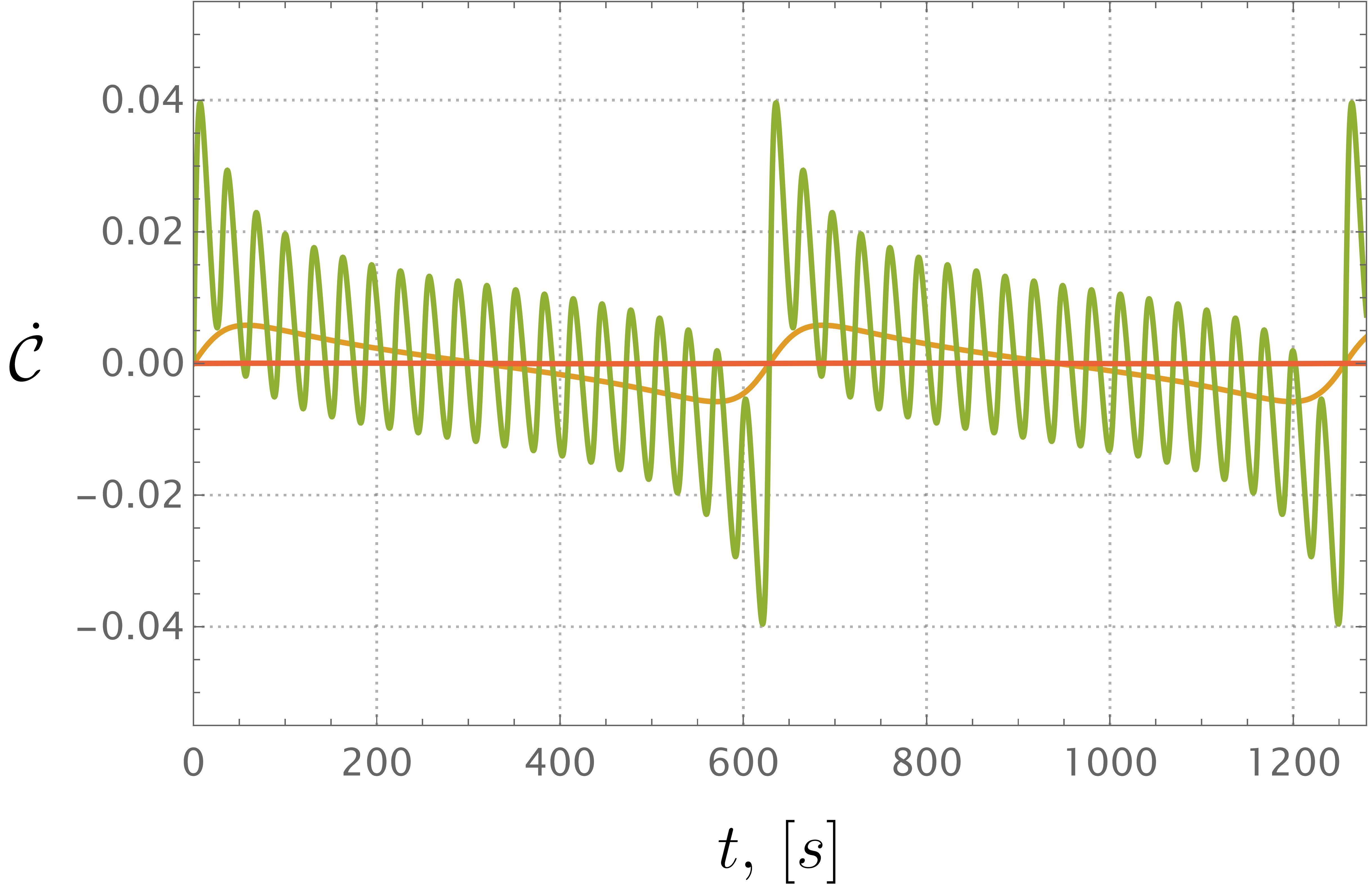}\vspace{-0.05cm}
\caption{Rate of complexity $\dot{\mathcal C}(t)$ at strong $B$. }\label{w1>>w2rate}
\end{subfigure}
 \hspace{0.7 cm}
\caption{ Case I: $\omega_{1,2}<\omega_{R1,R2}=1$. \textbf{(a)} Complexity $\mathcal C(t)$ at strong external magnetic filed $\omega_1\gg \omega_2$. We choose: $\omega_1=0.1,\, \omega_2=0.005$ ($\omega_c=0.095,\, \omega_0=0.022$).
\textbf{(b)} The corresponding rate of  complexity.
}\label{figEnsembles}
\end{figure}

\begin{figure}[H]
\centering \hspace{-1.0cm}
\begin{subfigure}{0.4\textwidth}
\includegraphics[width=8.3cm,height=5.55cm]{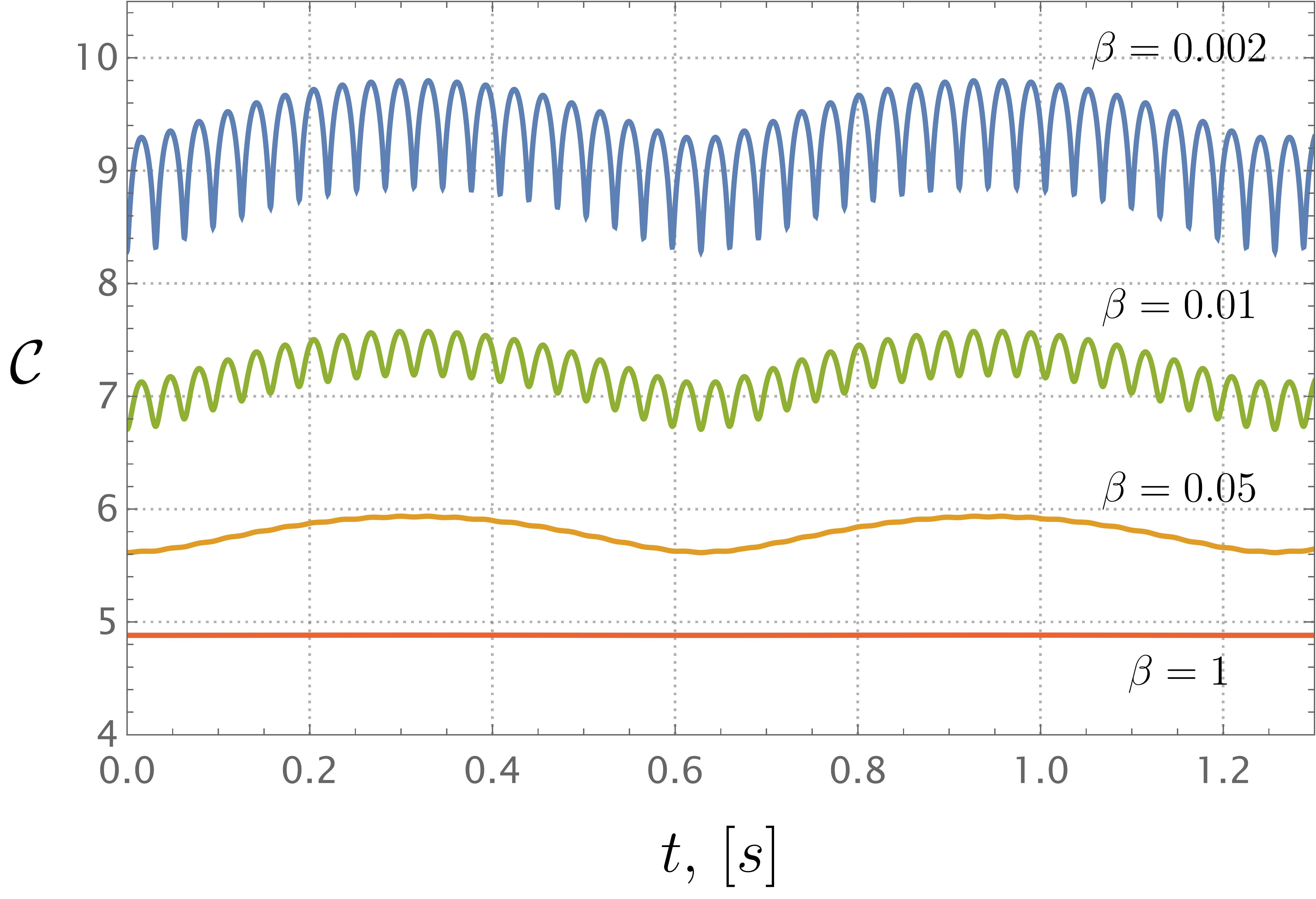}
\caption{Complexity $\mathcal C(t)$ at strong $B$.}\label{w1>>w2c}
\end{subfigure}
\hspace{1.5 cm}
\begin{subfigure}{0.4\textwidth}
\includegraphics[width=8.3cm,height=5.7cm]{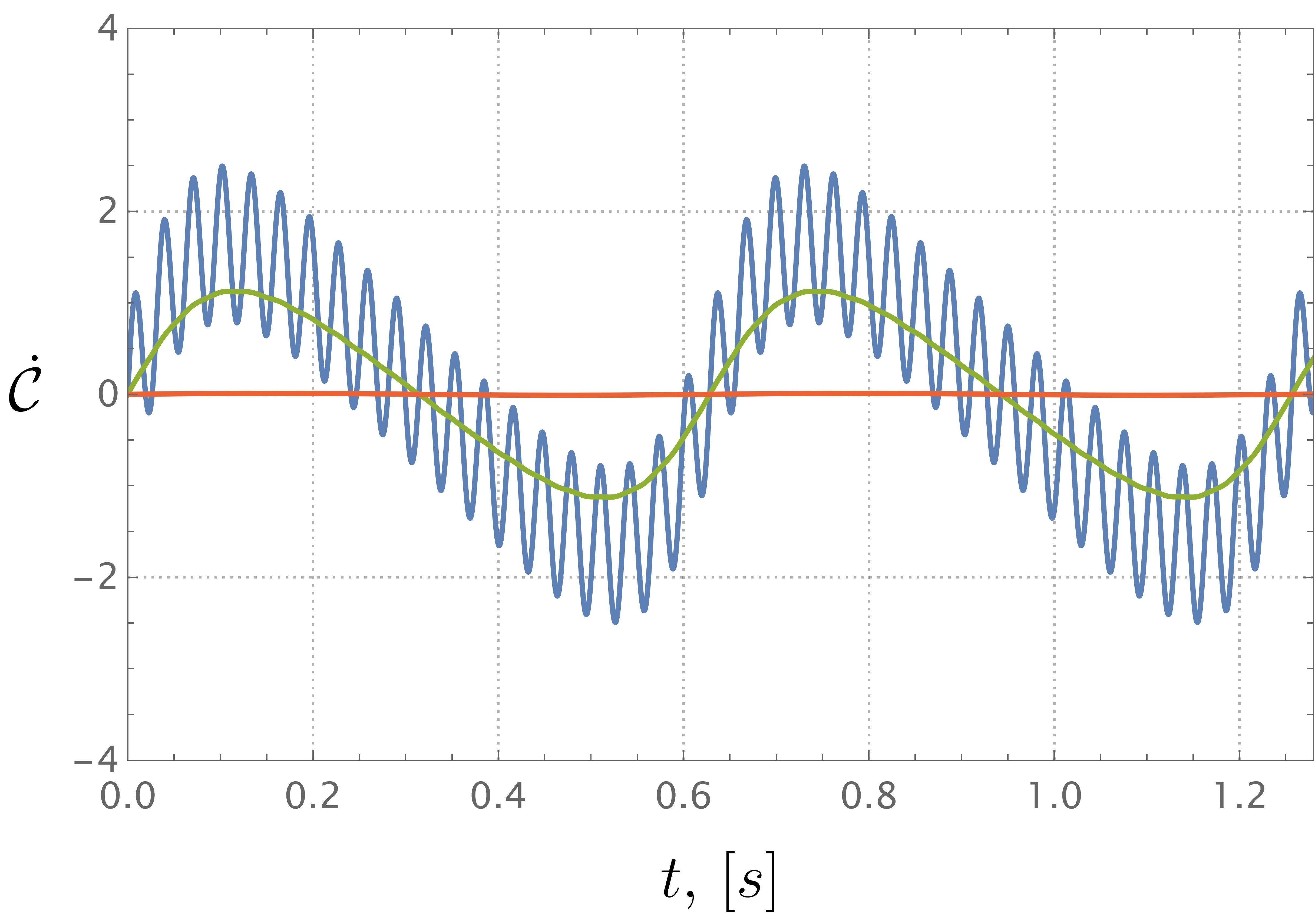}\vspace{-0.05cm}
\caption{Rate of complexity $\dot{\mathcal C}(t)$ at strong $B$. }\label{w1>>w2ratec}
\end{subfigure}
 \hspace{0.7 cm}
\caption{ Case II: $\omega_{1,2}>\omega_{R1,R2}=1$.\textbf{(a)} Complexity $\mathcal C(t)$ at strong external magnetic filed $\omega_1\gg \omega_2$. We choose: $\omega_1=100,\, \omega_2=5$ ($\omega_c=95,\, \omega_0=22$).
\textbf{(b)} The corresponding rate of  complexity $\dot{\mathcal C}(t)$, where the blue curve corresponds to $\beta=0.05$, green: $\beta=0.1$, and red: $\beta=1$.
}\label{figEnsemblesc}
\end{figure}

\noindent\textbf{Strong magnetic field}: In this case $\omega_c \gg \omega_0$, which results in $\omega_1 \gg \omega_2$ (see Eq. \eqref{frequencies}). This scenario is illustrated in Figs.\,\ref{w1>>w2} and \ref{w1>>w2c} for different values of the temperature $\beta$. The plots in Fig.\,\ref{w1>>w2} show complexity for $\omega_{1,2} < 1$, while Fig.\,\ref{w1>>w2c} corresponds to $\omega_{1,2} > 1$. Complexity is a periodic function with period dominated by the lower frequency $\omega_2$:  
\begin{equation}\label{Tw2}
\mathcal{T}\approx \frac{\pi}{\omega_2}.
\end{equation}
Furthermore, as the temperature increases, the oscillations in complexity with respect to $\omega_{1,2}$ become more pronounced, as demonstrated by the blue and green curves. Conversely, when the temperature decreases, these oscillations become more subdued, as shown by the orange and red curves. The red line, in particular, indicates the lower bound from Eq. \eqref{red}.

\begin{figure}[H]
\centering \hspace{-1.0cm}
\begin{subfigure}{0.4\textwidth}
\includegraphics[width=8.3cm,height=5.55cm]{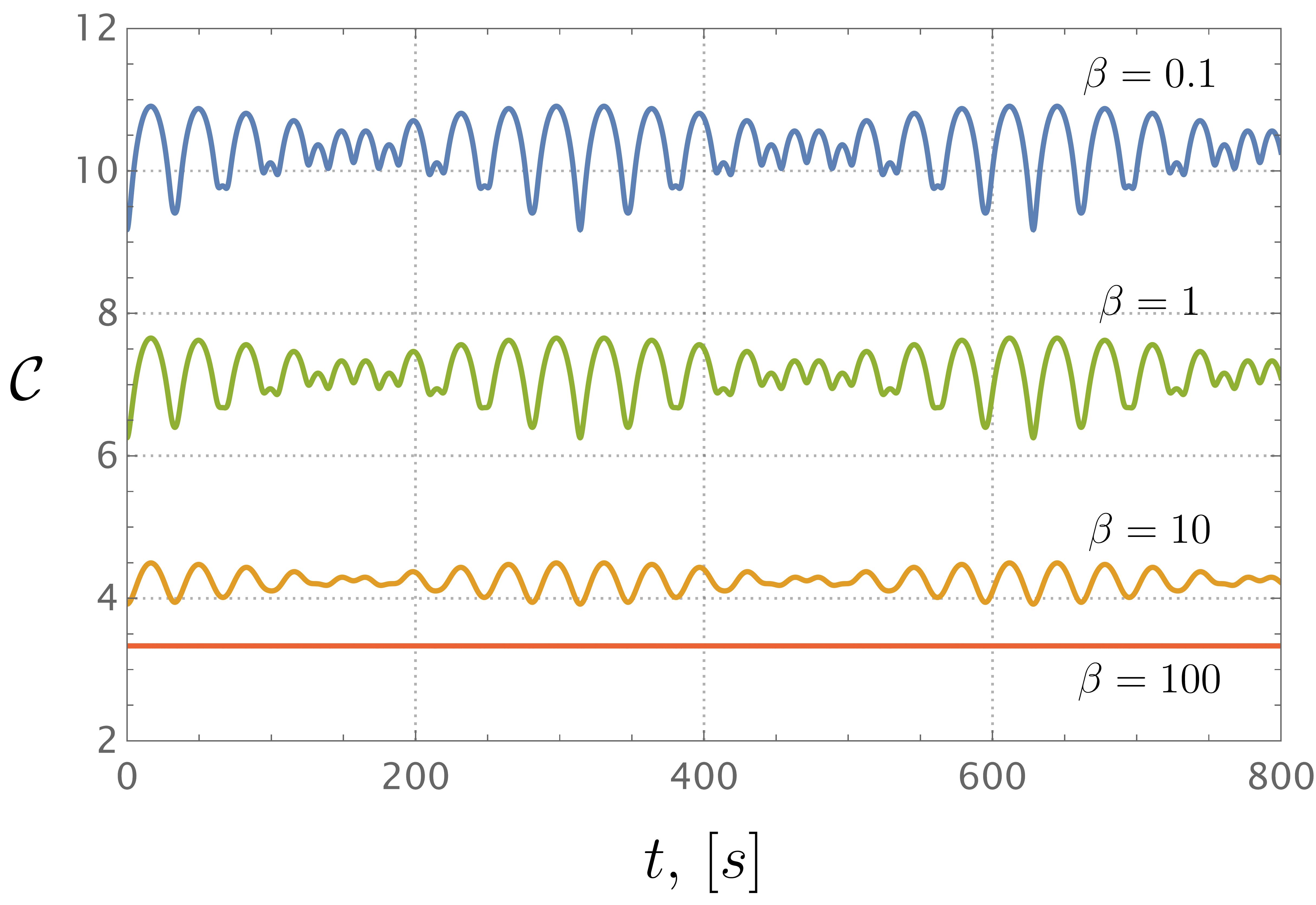}
\caption{Complexity $\mathcal C(t)$ at weak $B$.}\label{w1_w2}
\end{subfigure}
\hspace{1.5 cm}
\begin{subfigure}{0.4\textwidth}
\includegraphics[width=8.5cm,height=5.6cm]{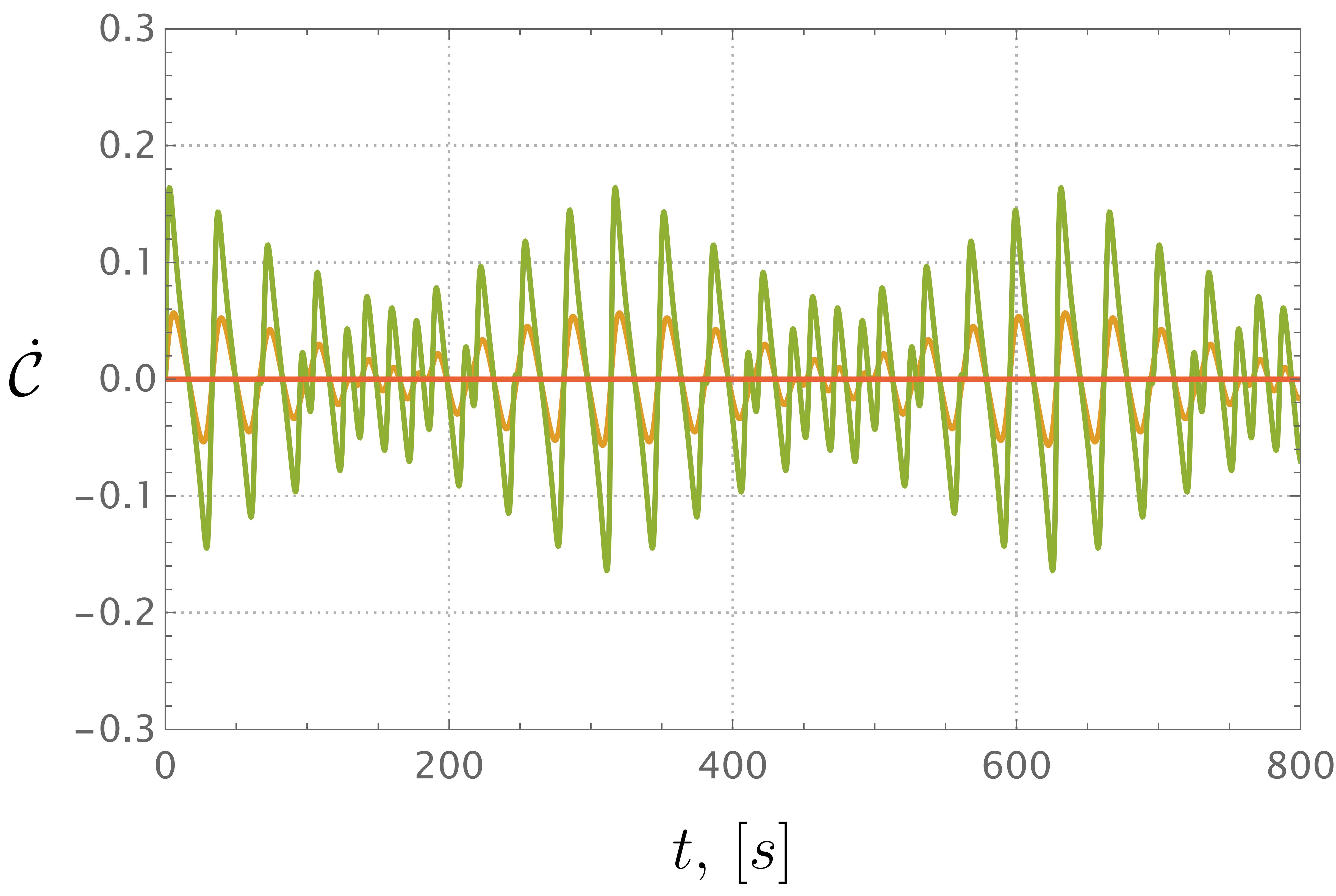}\vspace{-0.05cm}
\caption{Rate of complexity $\dot{\mathcal C}(t)$ at weak $B$.}\label{w1_w2_rate}
\end{subfigure}
 \hspace{0.7 cm}
\caption{ \textbf{(a)} Complexity $\mathcal C(t)$ under weak external magnetic filed $\omega_1\approx \omega_2$. We choose: $\omega_1=0.1,\, \omega_2=0.09$ ($\omega_c=0.01,\, \omega_0= 0.095$).
\textbf{(b)} The corresponding rate of complexity $\dot{\mathcal C}(t)$.}\label{w1appw2}
\end{figure}

\noindent \textbf{Weak magnetic field}: In this case $\omega_c \ll \omega_0$, resulting in $\omega_1 \approx \omega_2$. This situation is illustrated in Fig.\,\ref{w1appw2} for $\omega_{1,2}<1$, where we observe a beating effect with a period
\begin{equation} \label{periods}
\mathcal{T}\approx\frac{\pi}{\omega_1 -\omega_2}.
\end{equation}
This effect is more pronounced at high temperature (shown by the blue and green curves) and becomes less noticeable at low temperature (depicted by the orange and red curves). The complexity value corresponding to the red line once again represents the lower bound \eqref{red}. In the case of $\omega_{1,2}>1$ the behaviour of complexity is qualitative the same.

\begin{figure}[H]
\centering \hspace{-1.0cm}
\begin{subfigure}{0.4\textwidth}
\includegraphics[width=8.3cm,height=5.55cm]{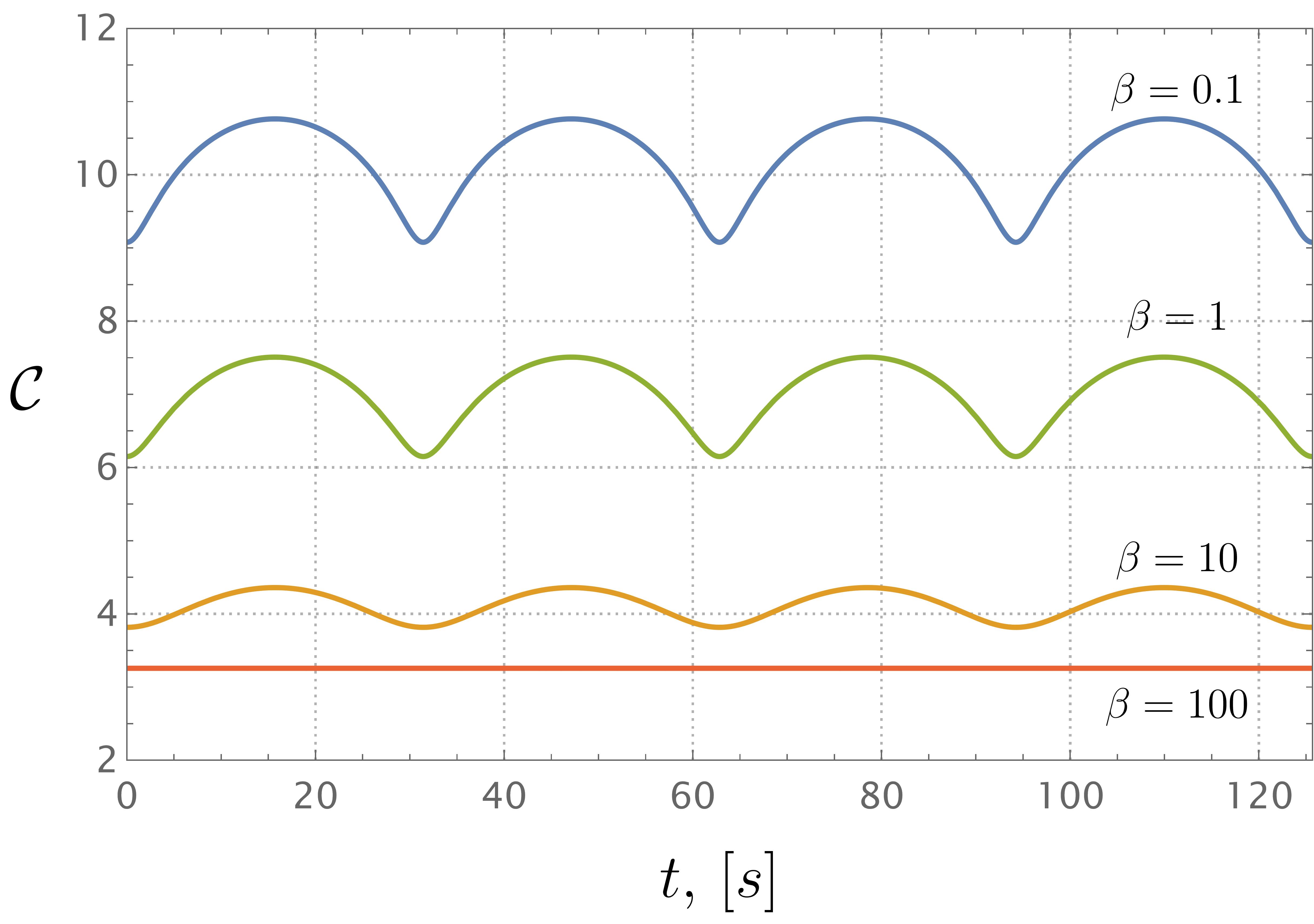}
\caption{Complexity $\mathcal C(t)$ at  $B=0$.}\label{w1_w2_0}
\end{subfigure}
\hspace{1.5 cm}
\begin{subfigure}{0.4\textwidth}
\includegraphics[width=8.5cm,height=5.45cm]{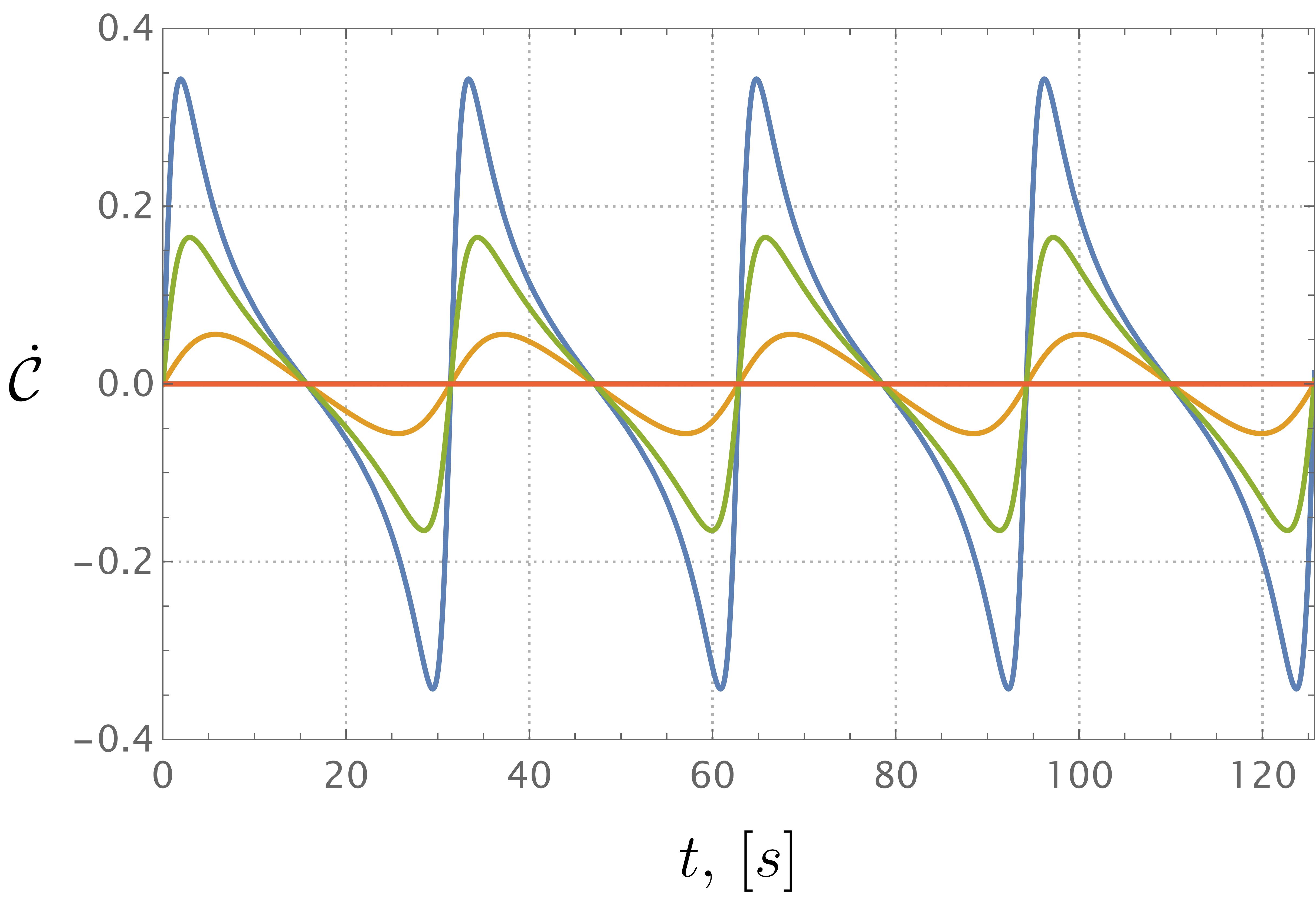}\vspace{-0.05cm}
\caption{Rate of complexity $\dot{\mathcal C}(t)$ at $B=0$.}\label{w1_w2_rate_0}
\end{subfigure}
 \hspace{0.7 cm}
\caption{ For $B=0$ one has $\omega_{1}=\omega_{2}=\omega_0$. We choose $\omega_0=0.1$ for demonstration. \textbf{(a)} Complexity $\mathcal C(t)$ without magnetic field.
\textbf{(b)} The corresponding rate of complexity $\dot{\mathcal C}(t)$. Here the blue curve correspond to the limit $\beta\to 0$ from \eqref{eqFRCBZ}.} \label{w1appw2_0}
\end{figure}

\noindent \textbf{Zero magnetic field:} Turning off the external magnetic field $\omega_c=0$ leads effectively to only one frequency $\omega_1=\omega_2=\omega_0$. The latter corresponds to two decoupled harmonic oscillators with equal frequencies. In this case, the time evolution of complexity is show on Fig.\,\ref{w1_w2_0}. The period of the oscillations is given by $\mathcal{T}=\pi/\omega_0$.
\noindent \textbf{Zero harmonic frequency:} The harmonic frequency $\omega_0=0$ can be excluded, resulting in $\omega_1=\omega_c$ and $\omega_2=0$. However, this exclusion introduces a nontrivial complication, as the calculation of the TFD states now demands a specific regularization procedure. This issue has been addressed in \cite{Radomirov:2024rjq}.

\section{Rate of complexity and Lloyd's bound} \label{lloyd_sect}

\subsection{Internal energy}

The internal energy of the TFD state follows directly from the partition function \eqref{PartF}: 
\begin{equation} \label{lloyd_bound}
U=-\,\frac{\partial \ln Z}{\partial \beta} =\frac{\hbar}{2} \bigg( \omega_1 \coth\frac{\beta\hbar\omega_1}{2} +\omega_2 \coth\frac{\beta\hbar\omega_2}{2} \bigg). 
\end{equation}
At very low temperature $(\beta\to\infty)$ the internal energy reduces to the energy of the ground state \eqref{energy}, which is a sum of the energies of two harmonic oscillators:
\begin{equation}\label{E00}
\lim_{\beta\to \infty} U=\frac{\hbar}{2}(\omega_1 +\omega_2)=E_{0,0}.
\end{equation}
At high temperature $(\beta\hbar\omega_i \ll 1)$ the quantum effects become negligible and the internal energy scales as a linear function of temperature
\begin{equation}\label{Uto0}
U\approx \frac{2}{\beta}= 2kT.
\end{equation}

With the internal energy known, we can now calculate the system's rate of complexity. As stated in \cite{lloyd}, the rate of complexity is constrained by the energy-related limit known as the Lloyd bound. In the following, we show that the quantum harmonic oscillator adheres to this bound.

\subsection{Rate of complexity}

The rate of complexity can be calculated form \eqref{Complexity} yielding
\begin{equation}
\dot{\mathcal C}(t) =\frac{1}{4\mathcal C(t)} \sum_{s=1}^8 \frac{\dot e_s}{e_s} \ln e_s. 
\end{equation}
At the zero-temperature limit the rate of complexity vanishes $\lim\limits_{\beta\to \infty} \mathcal{\dot{C}}(t) = 0$, which is represented by the red line on Figs.\,\ref{w1>>w2rate}-\ref{w1_w2_rate_0}. In contrast, at the high-temperature limit, the rate of complexity becomes
\begin{equation}\label{eqFRCBZ}
\lim_{\beta\to 0}  \dot{\mathcal C}(t) =\frac{1}{2\sqrt{2}} \sum_{i=1}^2 \frac{\omega_i (\omega_{Ri}^2-\omega_i^2)^2 \sin 2\omega_i t}{(\omega_{Ri}^2+\omega_i^2)^2 -(\omega_{Ri}^2-\omega_i^2)^2 \cos^2\omega_i t},
\end{equation}
which is a function with a finite amplitude. This is illustrated by the blue curve in Fig. \ref{w1_w2_rate_0}, where $\omega_1=\omega_2=\omega_0$.

\subsection{Lloyd bound}

\begin{figure}[H]
\centering \hspace{-1.0cm}
\begin{subfigure}{0.42\textwidth}
\includegraphics[width=8.3cm,height=5.55cm]{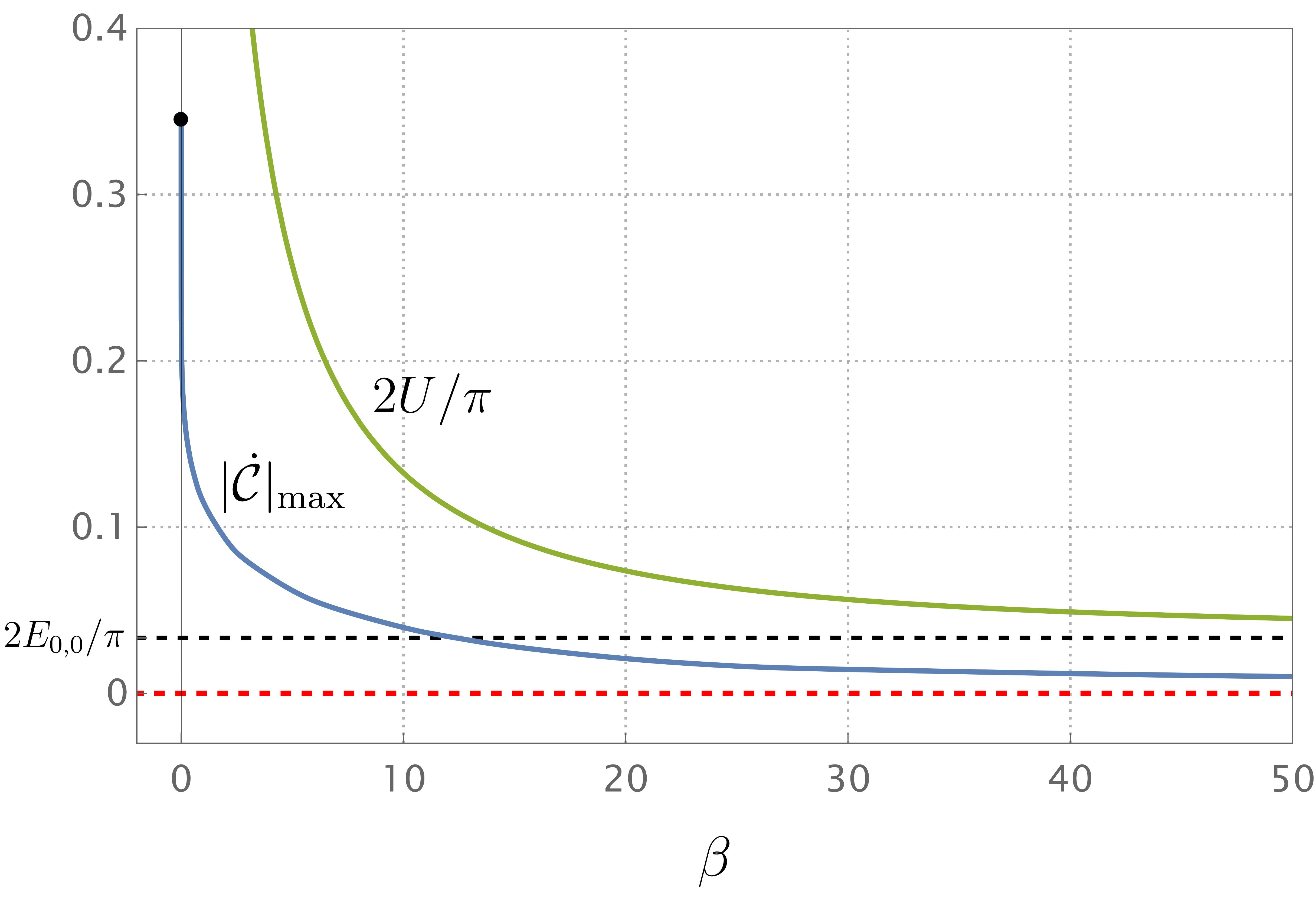}
\caption{$|\dot{\mathcal C}|_{\rm max}$ at strong $B$.}\label{w1_w2A1}
\end{subfigure}
\hspace{1.5 cm}
\begin{subfigure}{0.42\textwidth}
\includegraphics[width=8.3cm,height=5.55cm]{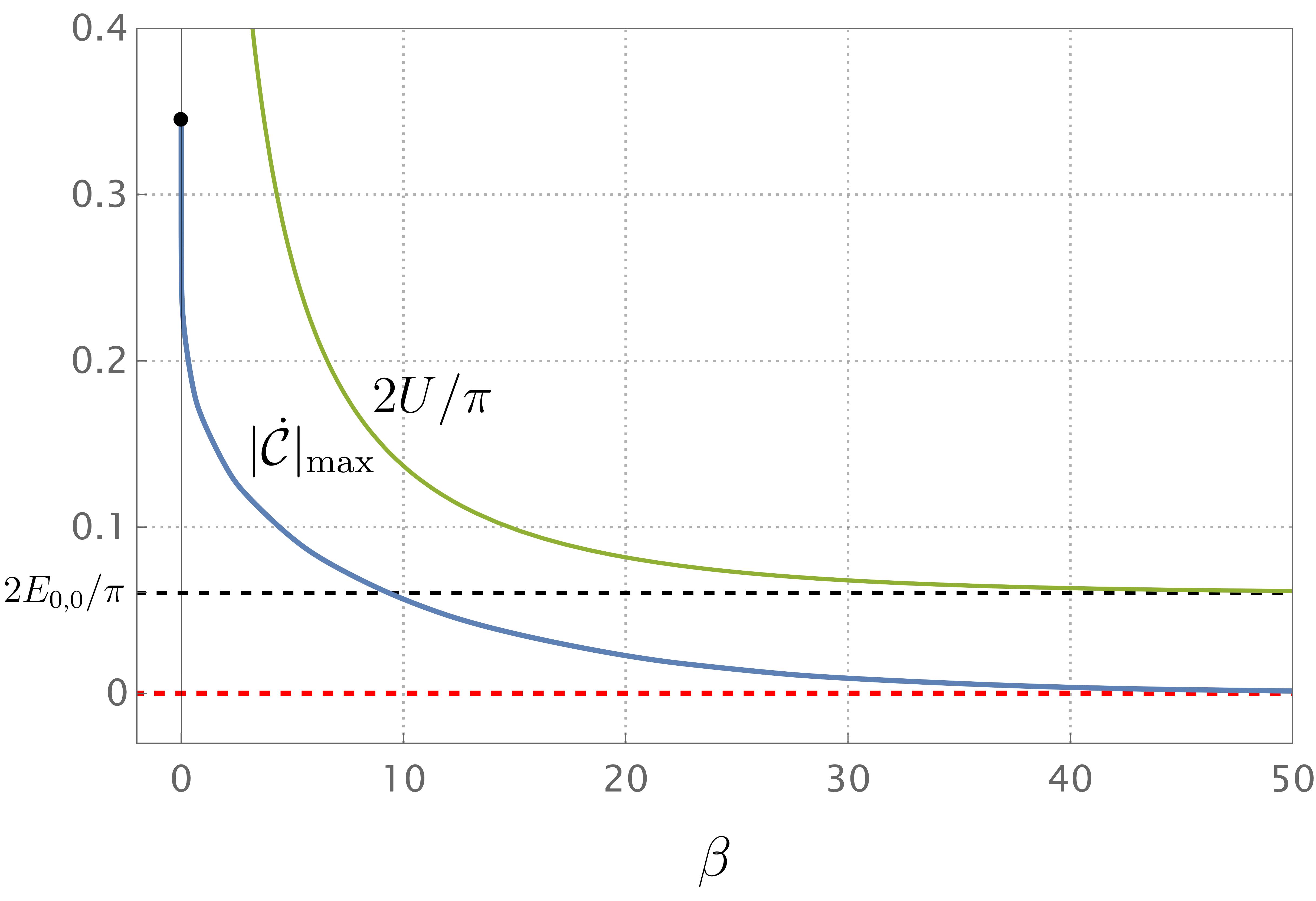}\vspace{-0.05cm}
\caption{$|\dot{\mathcal C}|_{\rm max}$ at weak $B$.}\label{w1_w2_rateA1}
\end{subfigure}
 \hspace{0.7 cm}
\caption{Maximum rate of complexity $|\dot{\mathcal C}|_{\rm max}$ blue and internal energy green as functions of $\beta$. \textbf{(a)} At strong magnetic field $\omega_1 \gg \omega_2$. We choose: $\omega_1=0.1,\, \omega_2=0.005$.
\textbf{(b)} At weak magnetic field $\omega_1 \approx \omega_2$. We choose: $\omega_1=0.1,\, \omega_2=0.09$. In both cases we set $\omega_{R1}=\omega_{R2}=1$ and $\hbar=1$.}\label{w1appw21}
\end{figure}

Finally, our goal is to compare the rate of complexity to the system's internal energy, known as the Lloyd bound, which according to quantum information theory, is given by \cite{lloyd}:
\begin{equation}\label{Lloyd}
|\dot{\mathcal C}|_{\rm{max}} \leq \frac{2U}{\pi\hbar}.
\end{equation}
In order to investigate whether the Lloyd bound is satisfied, we need to consider the maximum rate of complexity \( |\mathcal{\dot{C}}|_{\rm{max}} \), which we define as the global maximum of this quantity within every period of oscillation. We can demonstrate that the bound \eqref{Lloyd} is satisfied for all range of temperature as shown in Fig.\,\ref{w1appw21}. In this scenario, \( |\mathcal{\dot{C}}|_{\rm{max}} \) (represented by the blue curve) consistently remains below \(2U/\pi\) (depicted by the green curve). At low temperature, the green curve nears its minimum, proportional to the ground state energy (black dashed line), while the blue curve approaches zero (red dashed line). At high temperatures, the green curve diverges, while the blue curve approaches the black dot, indicating the upper limit of the rate of complexity. This upper limit corresponds to the amplitude of \eqref{eqFRCBZ}.

It is important to note that the ground state energies \eqref{E00} differ the two magnetic regimes. In any case, the ground state energy under a weak \(B\) field is always higher than that in a strong \(B\) field. Additionally, as \(\beta\) increases, the internal energy and the maximum rate of complexity approach their corresponding asymptotes much faster in the weak \(B\)-field regime Fig.\,\ref{w1_w2_rateA1}. For zero magnetic field $(\omega_1=\omega_2=\omega_0)$ one has twice the internal energy of harmonic oscillator
\begin{equation}
U=\hbar  \omega_0 \coth \frac{\beta  \hbar \omega_0}{2}.
\end{equation}
Even in this case the Lloyd bound still holds.

\section{Conclusion}\label{concl}

In this work, we explored the Nielsen complexity of a thermofield double state for a harmonic oscillator subjected to an external magnetic field. This was accomplished using the covariance matrix method as outlined in \cite{chapman2019complexity}, which is particularly well-suited for analyzing the evolution of Gaussian states. In general, Nielsen complexity defines the optimal paths connecting two such states. Building on these ideas, we obtained explicit results for the complexity and its rate, extensively studying the effects of temperature and the magnetic field.

We investigated the temperature dependence of complexity in two limiting cases: zero temperature and high temperature. In the zero temperature limit, the complexity saturates to a minimum positive value \eqref{red}. In the high temperature limit, the complexity becomes unbounded from above \eqref{C_Tinf}. In all scenarios, an increase in temperature leads to a growth in the system's complexity.

Regarding the effect of the magnetic field, we demonstrated that Nielsen complexity exhibits intriguing characteristics. For a strong magnetic field, complexity shows periodic behavior with a period primarily influenced by the lower frequency \eqref{Tw2}. For a weak magnetic field, complexity displays a beating effect (Fig.\,\ref{w1_w2}) with a period defined in \eqref{periods}. Furthermore, we showed that in the absence of a magnetic field, the complexity reduces to that of two non-interacting harmonic oscillators. The case of zero harmonic frequency has been already considered in \cite{}.

Finally, we analyzed the Lloyd bound of the system, which relates the system's internal energy to the maximum rate of complexity. We demonstrated that, regardless of temperature or magnetic field, the rate of complexity always adheres to this bound.

While this work focused on how an external magnetic field influences the complexity of a quantum oscillator, a similar study could be broadened to encompass more general systems. Such a comprehensive analysis may help clarify ambiguities in the definition of complexity across various contexts.

\section*{Acknowledgments}
The authors would like to express their gratitude to H. Dimov, A. Isaev and  S. Krivonos for their invaluable comments
and discussions. V. A. gratefully acknowledges the support by the SU grant 80-10-50/29.03.24,  
the Simons Foundation, the International Center for Mathematical Sciences (ICMS) in Sofia, SEENET-MTP and ICTP for the
various annual scientific events. M. R., R. R. and T. V. were fully financed by the European
Union-NextGeneration EU, through the National Recovery and Resilience Plan of the Republic
of Bulgaria, grant number BG-RRP-2.004-0008-C01.

\bibliographystyle{utphys}
\bibliography{ref}

\end{document}